\documentstyle[sprocl,epsfig]{article}
\def\Et{\ifmmode{E_T}\else{$E_T$\enskip}\fi}
\def\stop#1{\ifmmode{\tilde{t_{#1}}}\else{$\tilde{t_{#1}}$\enskip}\fi}
\def\D0{D\O}

\def\bbbar{$b \bar{b}$}
\def\ttbar{$t \bar{t}$}
\def\invpb{\ifmmode {{\rm \ pb}^{-1}} \else ${\rm pb}^{-1}$ \fi}
\def\invnb{\ifmmode {{\rm \ nb}^{-1}} \else ${\rm nb}^{-1}$ \fi}
\def\invfb{\ifmmode {{\rm \ fb}^{-1}} \else ${\rm fb}^{-1}$ \fi}
\def\Missing#1#2{{\mbox{$#1\kern-0.57em\raise0.19ex\hbox{/}_{#2}$}}}
\def\MEt{\Missing{E}{T}}

\def\zino#1{\ifmmode {\tilde \chi_{#1}^{0}} \else {$\tilde \chi_{#1}^{0}$} \fi}
\def\wino#1{\ifmmode {\tilde \chi_{#1}^{\pm}} \else {$\tilde \chi_{#1}^{\pm}$} \fi}
\def\squark{\ifmmode {\tilde q} \else {$\tilde q$} \fi}
\def\slepton{\ifmmode {\tilde l} \else {$\tilde l$} \fi}
\def\gluino{\ifmmode {\tilde g} \else {$\tilde g$} \fi}
\def\gravitino{\ifmmode {\tilde G} \else {$\tilde G$} \fi}

\def\be{\begin{equation}}
\def\ee{\end{equation}}
\def\bea{\begin{eqnarray}}
\def\eea{\end{eqnarray}}

\def\bprime{$b^\prime$}
\begin{document}
\title{SEARCHES FOR NEW PHYSICS AT THE TEVATRON}
\author{ K. WYATT MERRITT }
\address{Fermilab, P. O. Box 500, Batavia,\\ IL 60510, USA }
\maketitle\abstracts{
This paper summarizes searches at the Fermilab Tevatron
 for a wide variety of signatures
for physics beyond the Standard Model.
These include searches for supersymmetric particles, in the
two collider detectors and in one fixed target experiment. Also
covered are searches for leptoquarks, dijet resonances, heavy
gauge bosons, and particles from a fourth generation, as well
as searches for deviations from the Standard Model predictions
in dijet angular distributions, dilepton mass distributions, and
trilinear gauge boson couplings.}

\section{Introduction}
The wide variety of models which predict physics different from the
extremely successful
Standard Model can be divided into two types: theories which propose new
symmetries (such as supersymmetry and grand unified theories) and
theories which propose new dynamics (such as technicolor in all its
variations and theories of compositeness in the Standard Model's 
elementary building blocks).  Often the two types can predict the same
kind of new particle or effect (leptoquarks, for example, can appear in
both types, although perhaps with somewhat different phenomenology).
Most relevant for the current paper, however, is the observation that
almost all variations of such theories predict signatures for the new physics 
which are potentially within reach at the
center-of-mass energy of the current Tevatron.
This paper discusses the searches performed so far for such new physics, and
the prospects for future searches, at the Tevatron.

\section{ Searches for Supersymmetry}
\subsection{Phenomenology of Supersymmetry at the Tevatron}

Supersymmetry relates bosons to fermions, and its 
existence as a good symmetry at a high mass scale requires the existence of
a set of particles which are partners of their Standard Model counterparts,
with the same couplings but with different spin and R-parity. (R-parity is
a new multiplicative quantum number assigned so that the particles of the 
Standard Model have R-parity = $+1$, while their supersymmetric partners have
R-parity = $-1$.)  The breaking of supersymmetry at a scale below the 
unification scale implies that the masses of the superpartners
will be different (presumably, higher) than those of ordinary 
particles.  The list of particles and superparticles is given in
Table~\ref{SusyTable}.
The reader is referred to the reviews in Ref. 1 for 
discussion of theories of supersymmetry; we
discuss here the signatures of supersymmetry which can be seen at the
Tevatron.  

\begin{table*}[hbt]
\setlength{\tabcolsep}{1.5pc}
\caption{Supersymmetric particle spectrum}
\label{SusyTable}
\begin{tabular*}{\textwidth}{l|cl}
\hline
\multicolumn{1}{c} {R-parity = 1} & \multicolumn{2}{c} {R-parity = -1} \\
\hline
        q          &  $\squark_{R}, \squark_{L}$ & squarks\\
        l           &   \slepton & sleptons\\
        g          &   \gluino   & gluinos\\
                    &                  & \\
   $W^{\pm}$, $H^{\pm}$ &  \wino{1},\wino{2} & charginos\\
                    &                    & \\
   $Z^{0}$,$h^{0}$,$H^{0}$, $A^{0}$,$\gamma$  &
                 \zino{1},\zino{2}, \zino{3},\zino{4}    & neutralinos\\
                       &                     &  \\
   $G$             &   \gravitino  & gravitinos \\

\hline
\end{tabular*}
\end{table*}

All the supersymmetry searches reported to date from the Tevatron assume 
R-parity conservation, which implies that the SUSY particles are produced in
pairs, and decay in cascades which must terminate in two LSP's (LSP = lightest
supersymmetric particle).  In most of the searches, a SUSY model is chosen in
which the LSP is the lightest neutralino state \zino{1}.  The LSP 
is constrained to be neutral and non-interacting, so it 
produces a missing transverse energy (\MEt) signature in the detectors.   
The exact particle 
composition of the decay channels, and hence the branching fractions into
various final state topologies, are very dependent on the details of the
SUSY model.  In calculating acceptance for the various final states, the
experiments generally choose simplifying assumptions to limit the
model space covered.  The most popular simplifying assumptions are
derived from supergravity theories.\cite{SUGRAref}

The top squarks --- the supersymmetric partners of the top quark ---
are a special case for SUSY searches.
 The large mass of the top quark \cite{D0-CDFtop} has
the effect in SUSY models of producing different Yukawa
couplings for the top squarks which result in a pair of
mass eigenstates \stop{1}, \stop{2}, the lighter of which
can be lighter than the top quark itself.  The
decays of the \stop{1} would be top-like, either to
$W + b + LSP$ or to $\wino{1} + b$, {\em unless} these are
not kinematically allowed.  There is a region in the
$m_{\stop{1}}$ vs. $m_{LSP}$ plane, for \stop{1} lighter than
\wino{1},  where the only allowed decay
is $ \stop{1} \rightarrow c + LSP$, making searches simple
and model-independent (although not easy). 

We now turn to the specific SUSY searches which have been 
reported to date from the Tevatron.  There are searches for
squarks and gluinos (including one from the fixed target 
program), for charginos and neutralinos, for the lightest
top squark, and for the charged Higgs.  There is also an
investigation of the two-photon + \MEt\ final state, inspired
by models in which radiative decays of the neutralino
may play more of a role
than in the SUGRA-inspired models commonly used to
set limits in the other searches.   In some of these models,
the gravitino rather than the lightest neutralino is the LSP.  

\begin{figure*}[htb]
\begin{minipage}{0.48\linewidth}  
\begin{center}
\mbox{\mbox{\epsfig{figure=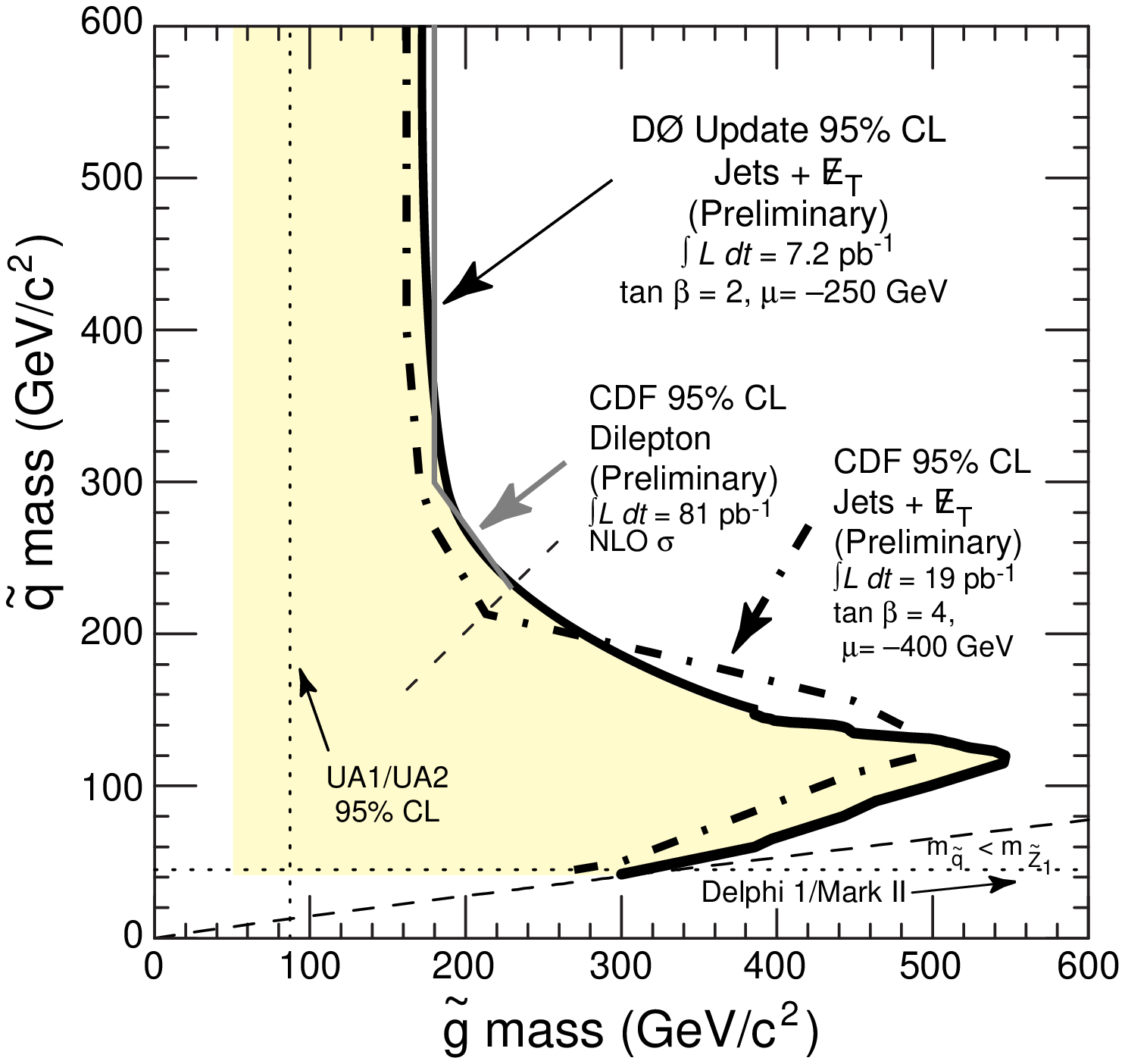,width=50mm}}}
\end{center}
\caption{The 95\% 
CL limits in the $m_{\squark}-m_{\gluino}$ plane from three
\squark-\gluino searches at CDF and \D0.}
\label{Combsqgl_fig}
\end{minipage}
\hfill
\begin{minipage}{0.48\linewidth}
\begin{center}
\mbox{\mbox{\epsfig{figure=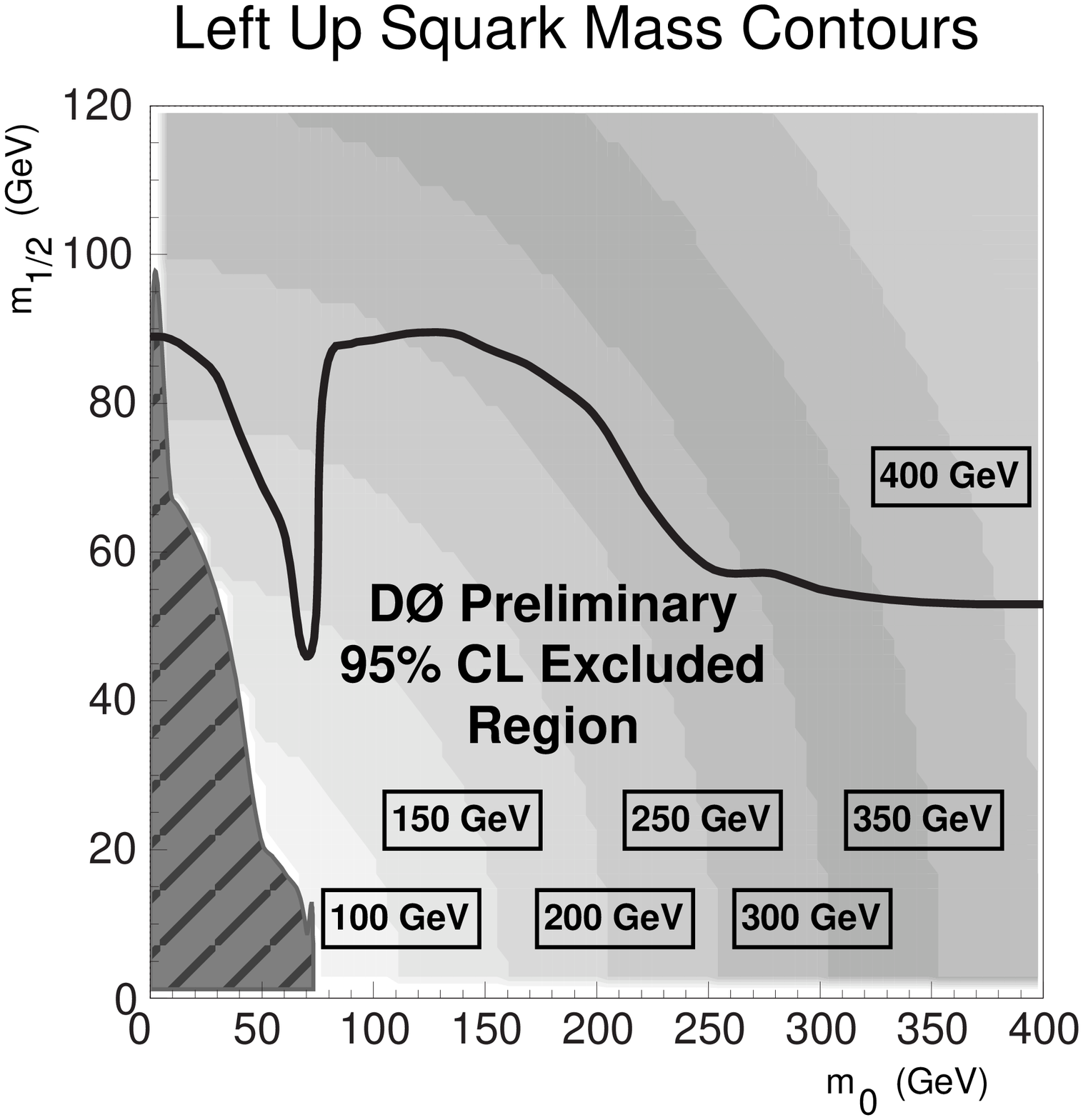,width=50mm}}}
\end{center}
\caption{The 95\% 
CL limits in the $m_{0}$ - $m_{1 \over 2}$ plane from the
\D0 dielectron \squark-\gluino search. The hatched region 
is excluded by theoretical constraints.}
\label{Sqgl_lep}
\end{minipage}
\end{figure*}

\subsection{Squarks and Gluinos}

Squarks and gluinos can be strongly pair-produced at the Tevatron
collider, with cross sections for \squark-$\widetilde{\overline{q}}$, \gluino-\gluino,
and \squark-\gluino determined in standard QCD calculations, 
depending on the masses of the new particles.  We expect the final states to consist
of either multijets + \MEt\ or jets + leptons + \MEt.  The \MEt\ would
be due to the existence of two relatively massive LSPs in the final 
state.   Four searches for such pair production have been completed.
Both \D0~\cite{D0sqgl} and CDF~\cite{CDFsqgl} have searched for the multijet + \MEt\
final state,
and the contours from these searches are shown in Fig. \ref{Combsqgl_fig}.
Both these searches use only the data from the 1992-93 Tevatron run,
and also use a leading order cross section calculation.
Also shown in that figure is the contour from a search by CDF 
\cite{CDFsqgl_lep} for dileptons
+ jets + \MEt, using the 1992-1995 data, and using an NLO cross section
calculation~\cite{NLOsqgl} which is $\approx$ 20~\%  larger than the LO
cross sections.  The reach achieved
in the dilepton search,
with the more favorable cross section and 
larger luminosity,  is comparable to the reach in the
hadronic channel over the range where the \squark mass 
is larger than the \gluino mass.

\D0  has conducted a search for dielectrons + jets + \MEt, using the 1994-1995
data, but it is reported within a different model framework than the searches 
above.\cite{D0_sqgl_lep}    This limit was calculated for a consistent supergravity model as a
function of $m_{0}$ and  $m_{1 \over 2}$, with the other SUGRA parameters
fixed to be tan $\beta$ = 2, $A_0$ = 0, and
sgn($\mu$) = -1, and the result is shown in Fig. ~\ref{Sqgl_lep}, in
which the limit curve in the  $m_{0}$ vs. $m_{1 \over 2}$ plane is compared
to the mass contours of the \squark.
The significant dip in the cross section
limit is due to the changing mass relationships --- specifically,
the limit worsens for $m_0$ below 100 GeV, 
when the $\tilde{\nu}$ becomes light enough to
provide an invisible decay mode for the \zino{2}, then recovers
as the $\tilde{e}$ becomes light enough to provide renewed dielecton
modes.

E761, a fixed target experiment designed to study hyperon decays, has
also made a search for the expected decays of supersymmetric baryons
which would exist in certain theories which call for extremely light
gluinos.\cite{Lightgluino}  The theoretical prediction calls for SUSY
baryons in a mass range of 1700-2500 MeV, with a lifetime of order
50-500 ps.  The limit obtained in the search \cite{E761} rules out
masses between $\approx$ 1700 and 2300 MeV, thus 
substantially limiting but not closing
the window available for these light gluino states.

\subsection{Charginos and Neutralinos}

\begin{figure*}[htb]
\begin{minipage}{0.48\linewidth}  
\begin{center}
\mbox{\mbox{\epsfig{figure=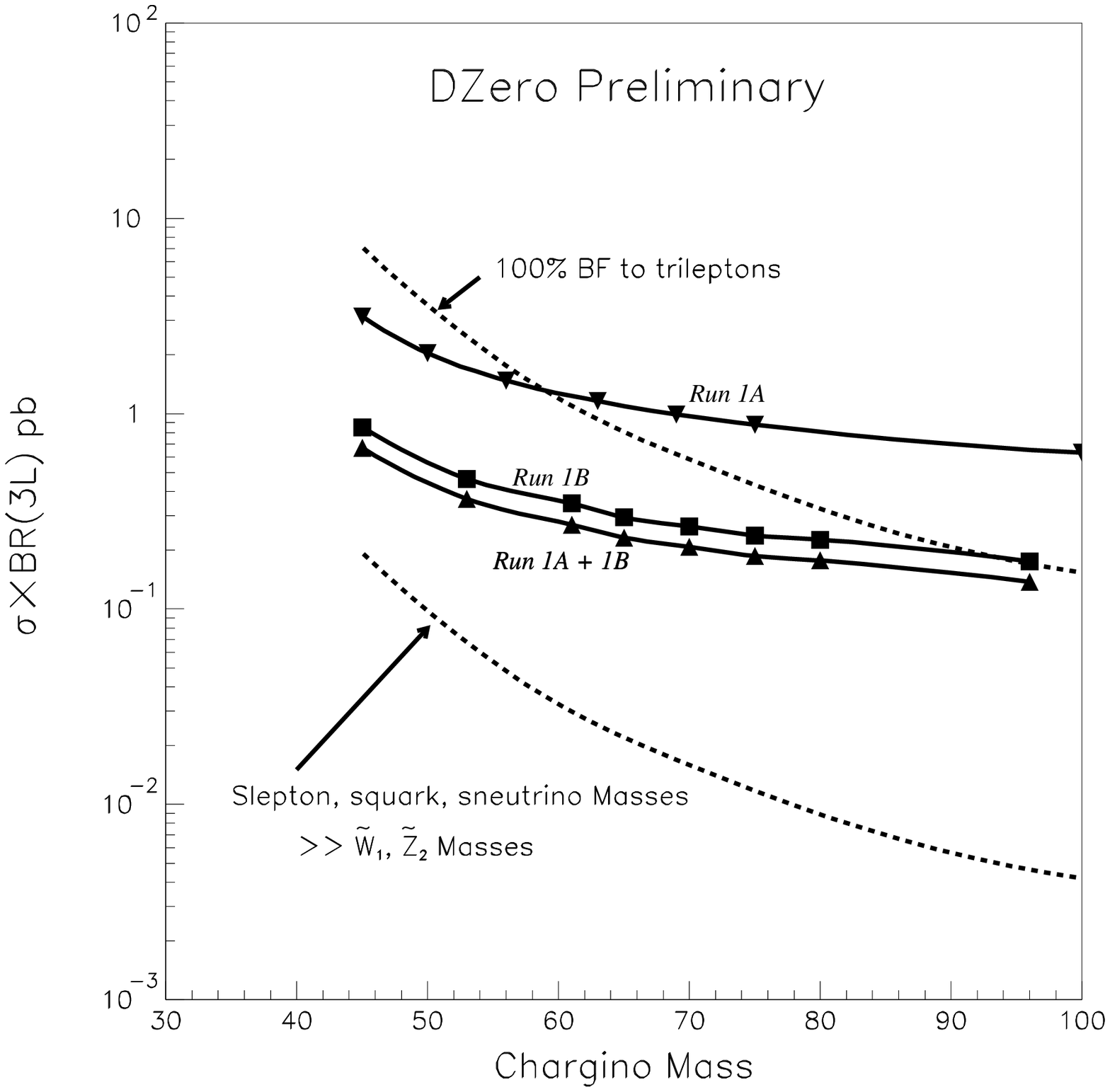,width=50mm}}}
\end{center}
\caption{ The 95\% 
CL limit on cross section $\times$ branching ratio into a single trilepton
channel vs. \wino{1} mass, from the \D0 trilepton search.}
\label{D0_WZ}
\end{minipage}
\hfill
\begin{minipage}{0.48\linewidth}
\begin{center}
\mbox{\mbox{\epsfig{figure=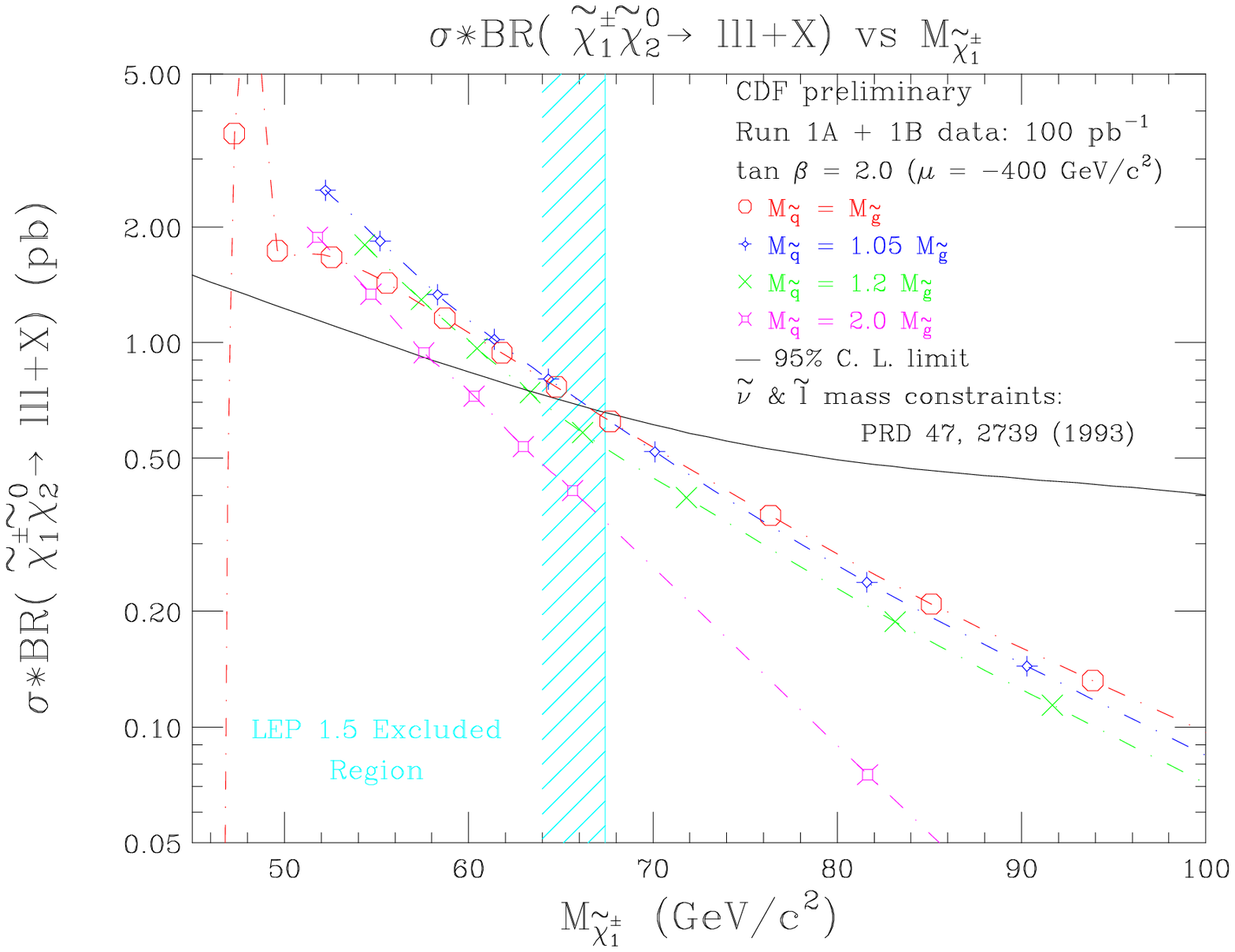,width=50mm}}}
\end{center}
\caption{The 95 \%
limit on cross section $\times$ branching ratio into the {\em four}
trilepton channels, from the CDF trilepton search.}
\label{CDF_WZ}
\end{minipage}
\end{figure*}

Two of the lowest mass chargino and neutralino
states of the Minimal Supersymmetric Standard Model can be weakly pair produced at the
Tevatron: $p\overline{p} \rightarrow
\wino{1} + \zino{2}$.  The chargino mass region starting 
just above the existing LEP I limit 
of approximately 45 GeV (which actually corresponds, in a
GUT-inspired
MSSM, to a reasonably high gluino mass) is accessible at the Tevatron.
The cleanest channels for such a search 
are
the purely leptonic decay channels, where the signature would be three leptons plus
\MEt.  The branching fraction for these channels depends sensitively 
on
the mass of sleptons in the theory, and other parameters.
Both \D0 and CDF have searched in four channels for this signature: $eee$, 
$ee\mu$,  $e\mu\mu$,
and $\mu\mu\mu$, using the 1994-95 data sample.\cite{D0_WZ,CDF_WZ}
The limit curves obtained are shown in Figs.~\ref{D0_WZ} and~\ref{CDF_WZ}.

\subsection{The Lightest Top Squark}

\begin{figure*}[htb]
\begin{minipage}{0.48\linewidth}  
\begin{center}
\mbox{\mbox{\epsfig{figure=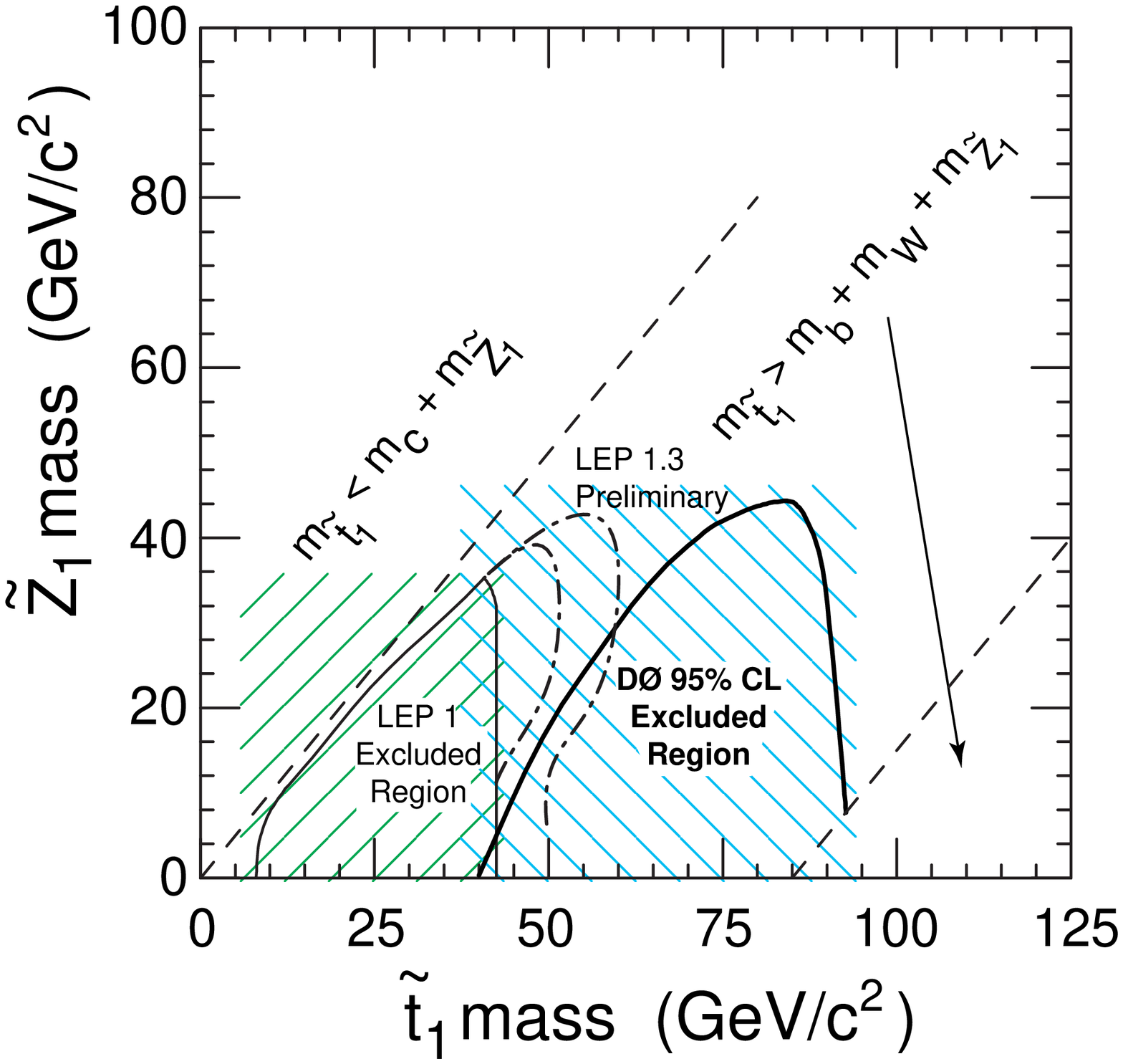,height=60mm}}}
\end{center}
\caption{ The 95\% 
CL limit in the $m_{\stop{1}}-m_{LSP}$ plane from 1992-93 data
from \D0.}
\label{D0stop}
\end{minipage}
\hfill
\begin{minipage}{0.48\linewidth}
\begin{center}
\mbox{\mbox{\epsfig{figure=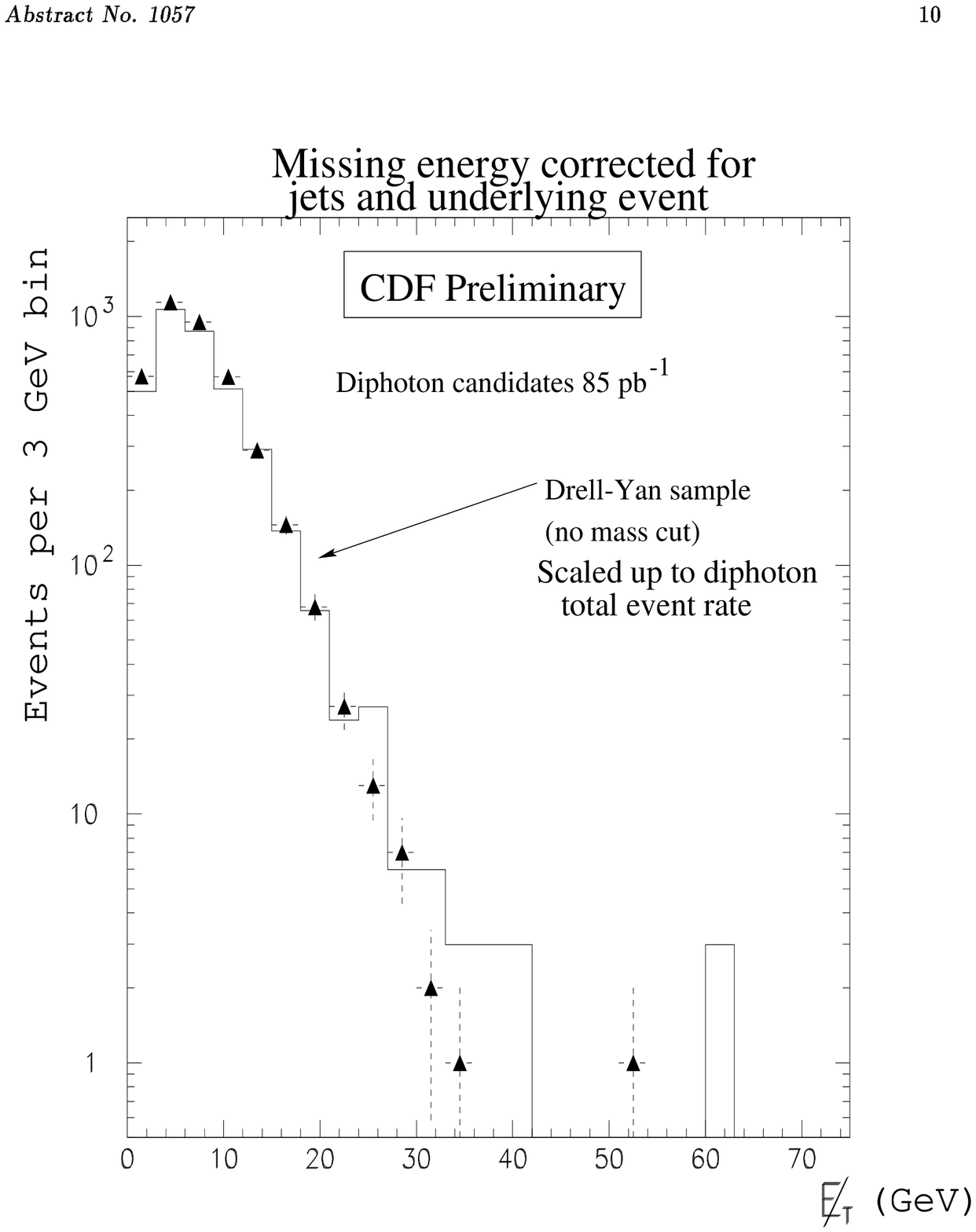,height=60mm}}}
\end{center}
\caption{The observed spectrum of diphoton invariant mass,  from
a CDF investigation of the channel diphoton + \MEt.}
\label{CDFdiphoton}
\end{minipage}
\end{figure*}

\D0 has searched in the 1992-93 data set for the lightest top squark
\stop{1}, in the topology of two acoplanar jets + \MEt.\cite{D0_stop}  As mentioned 
above, for certain mass regions of the various SUSY particles, this
channel is the only kinematically accessible one, and hence the
limit is completely model-independent given only that
$m_{\stop{1}} <  m_{W + b + LSP}$  {\em and} 
$m_{\stop{1}} <  m_{\wino{1} + b}$.  The limit is shown in 
Fig. \ref{D0stop}.

\subsection{The Charged Higgs}

CDF  reports a search for the charged Higgs boson in the framework
of minimal SUSY via its appearance in top decays: $ t \rightarrow
b + H^{+}$ followed by $H^{+}~\rightarrow~\tau~+~\nu_{\tau}$.\cite{CDF_Hplus}
This limit rules out a charged Higgs with a mass below 140 GeV for 
$tan~\beta~>~200 $.

\subsection{Radiative SUSY Decays}

Interest in models where radiative decays of SUSY particles are 
important~\cite{Kane_etal} was recently piqued by observation
of a single event at CDF with the topology of 2 electrons, 2 photons,
and large \MEt.   These models in turn predict SUSY signatures in
channels such as 2 photons + \MEt.  CDF showed an examination
of the 2 photon channel at this conference~\cite{CDF_diphoton} which
found no indication of an excess above background as shown in 
Fig. ~\ref{CDFdiphoton}.

\section{Searches for Dijet Resonances}

Many new states can have substantial branching fractions into
two jets, but this signature is difficult to observe inside the QCD
background.   The search limits obtained depend on assumptions
about the line width of the resonance being searched for. 

\subsection{Excited Quarks and Other Dijet States in Dijet Events}

\begin{figure*}[htb]
\begin{minipage}{0.48\linewidth}  
\begin{center}
\mbox{(a)\mbox{\epsfig{figure=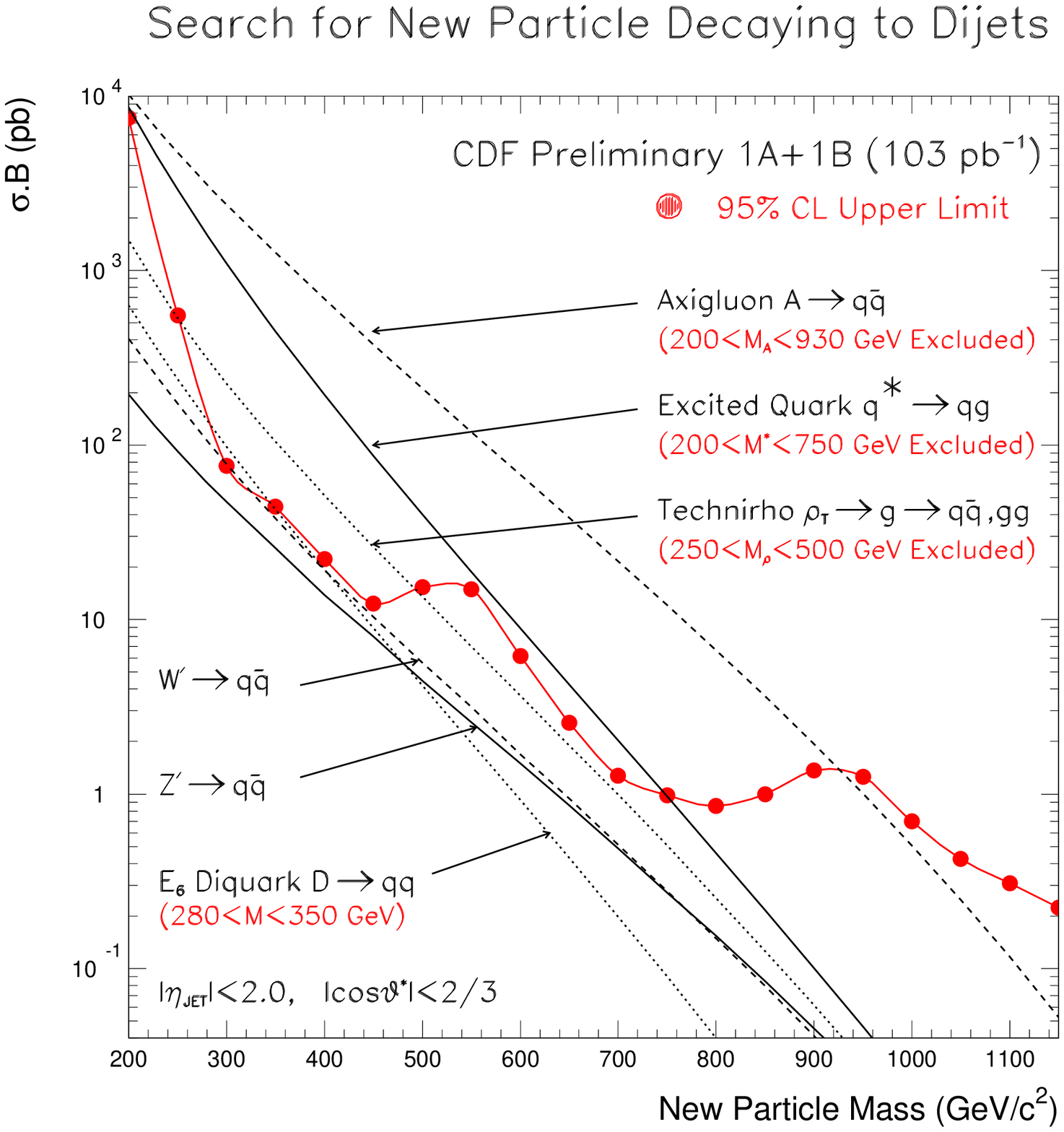,height=60mm}}}
\end{center}
\end{minipage}
\hfill
\begin{minipage}{0.48\linewidth}
\begin{center}
\mbox{(b)\mbox{\epsfig{figure=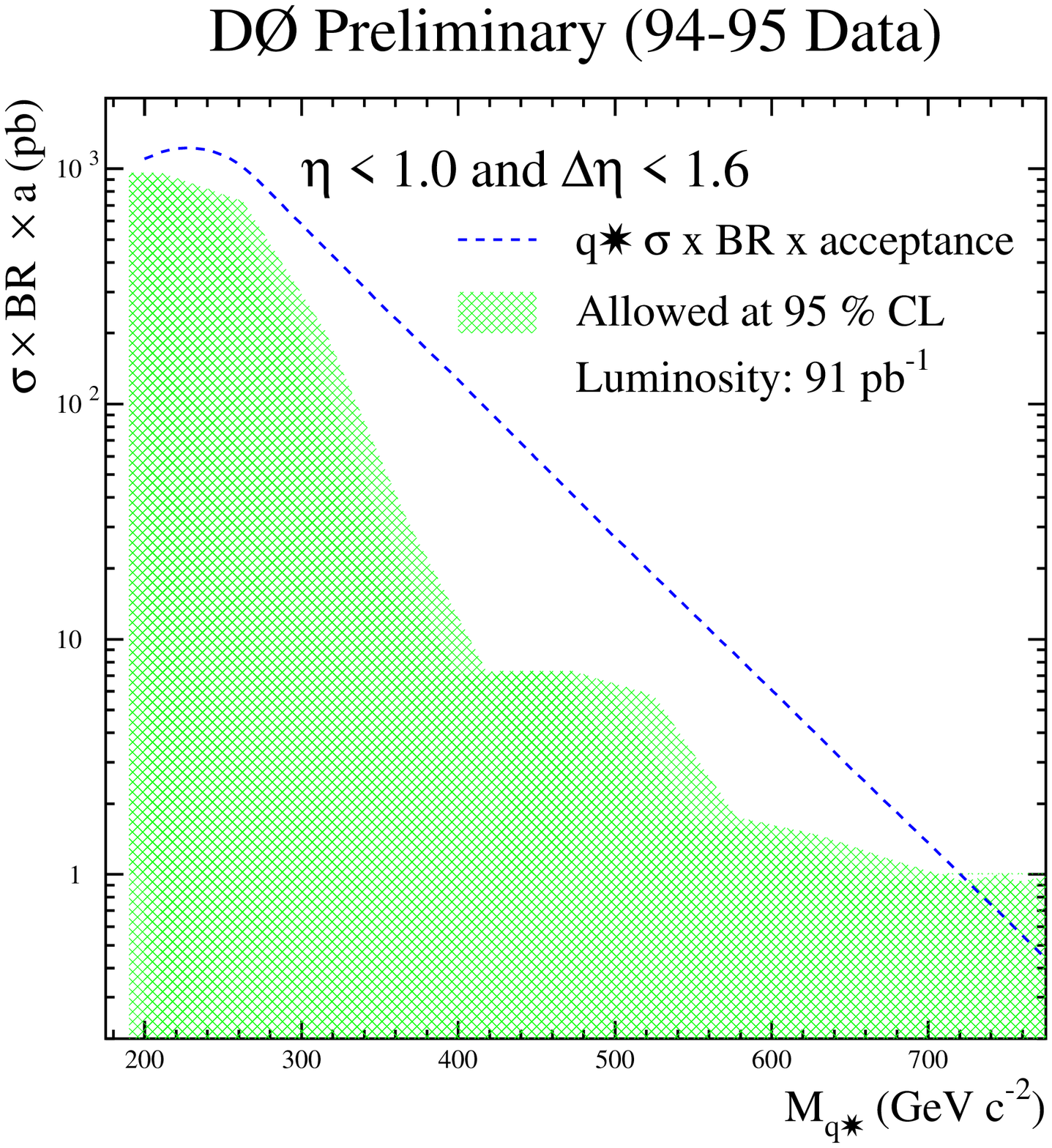,height=60mm}}}
\end{center}
\end{minipage}
\caption{(a) The 95\% 
limit on cross section $\times$ branching ratio into excited quarks,
compared to the theoretical excited quark cross section and cross
sections for some other new particles, from the CDF
dijet analysis.
(b) The 95 \%
limit on cross section $\times$ branching ratio into excited quarks,
compared to the theoretical excited quark cross section, from the \D0
dijet analysis.}
\label{Dijets}
\end{figure*}

From the dijet mass spectra obtained by CDF and \D0, cross section
upper limits can be obtained for the addition of a resonance of a
given width.  CDF~\cite{CDF_dijet} has reported from the 1992-95 data set a limit
curve, shown in Fig.~\ref{Dijets}(a).
\D0 reports at this conference~\cite{D0_dijet} a limit in the same channel from
1994-95 data shown in Fig.~\ref{Dijets}(b).

\subsection{Dijet Resonance Produced with a $W$ Boson}

\begin{figure*}[htb]
\begin{minipage}{0.48\linewidth}  
\begin{center}
\mbox{\mbox{\epsfig{figure=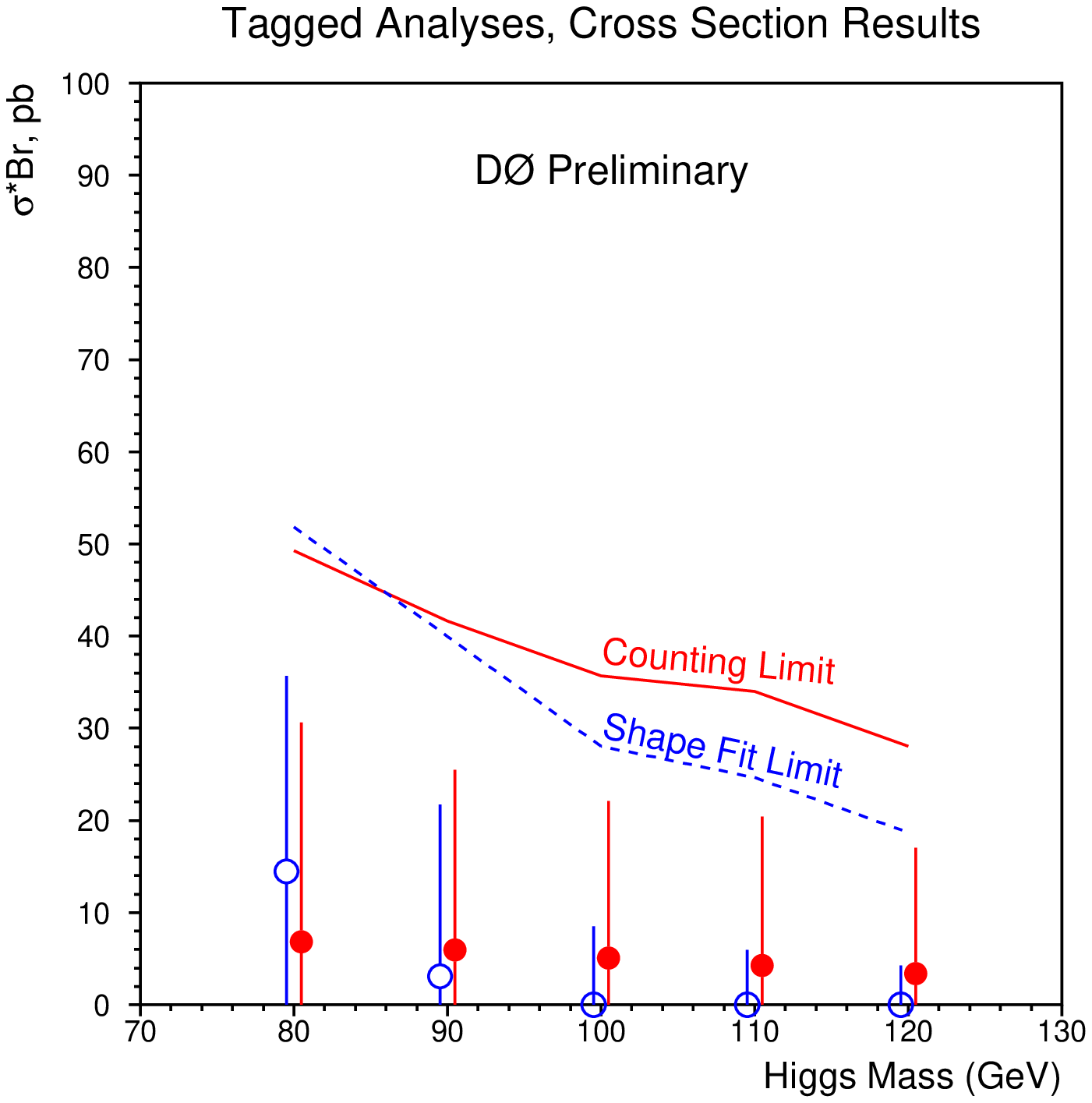,height=60mm}}}
\end{center}
\caption{The 95\%
CL cross section limits on production of a Higgs-like resonance
decaying to \bbbar, produced in association with a $W$ boson. }
\label{D0Higgs}
\end{minipage}
\hfill
\begin{minipage}{0.48\linewidth}
\begin{center}
\mbox{\mbox{\epsfig{figure=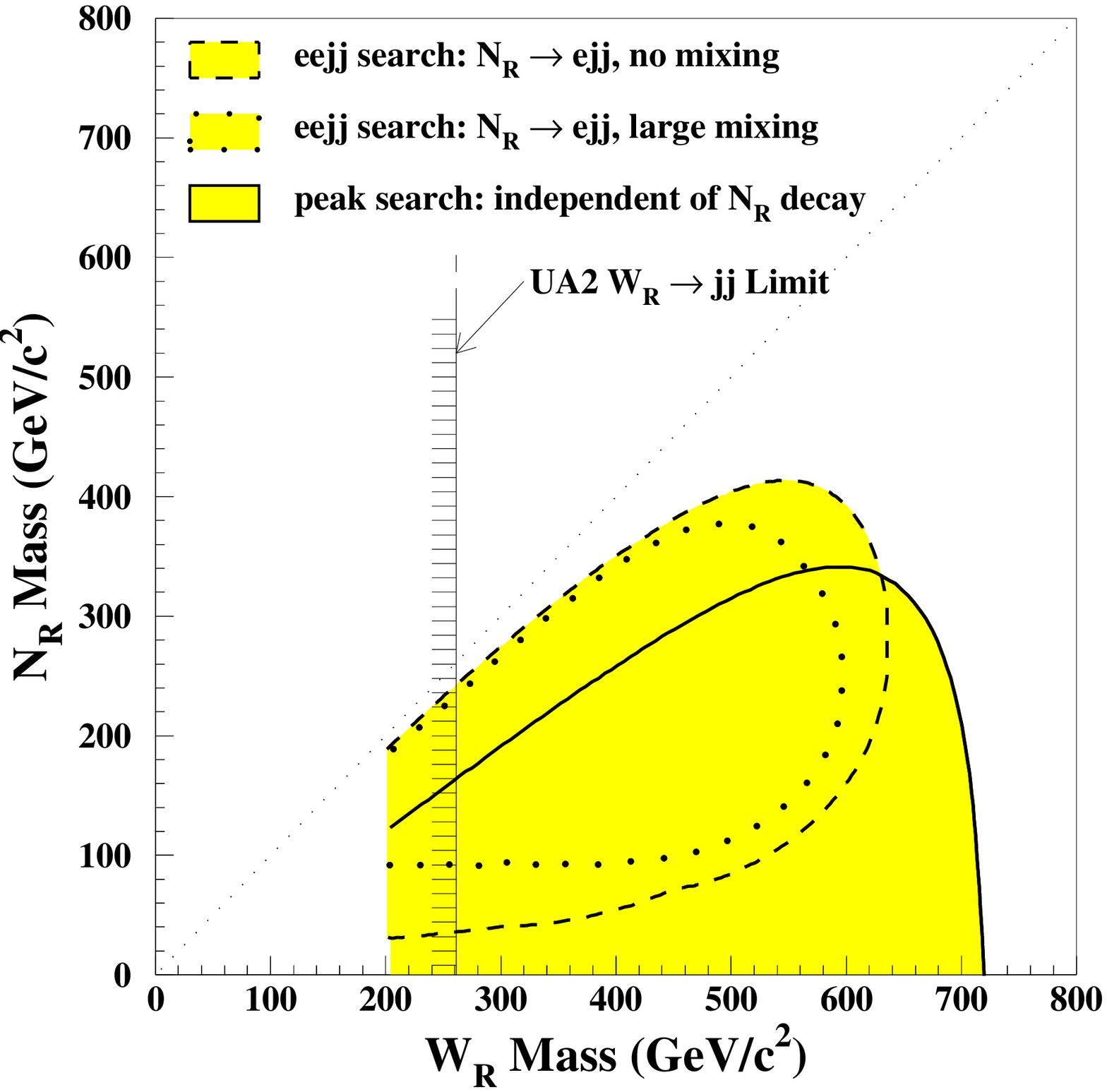,height=60mm}}}
\end{center}
\caption{
The limit contours from two different searches for a
heavy $W$ at \D0. See the text for a description of the searches.}
\label{D0Wright}
\end{minipage}
\end{figure*}

Some new particles (Standard Model or SUSY Higgs or technirhos, for
example) could be produced in association with $W$ bosons and then
decay to two jets.  The conventional Higgs cross section is lower than
the expected sensitivity in Run I, but the technirho cross 
section~\cite{Technicolor} 
is at least of the same order of magnitude.  \D0
reports at this conference a cross section times branching fraction
limit on such a resonance in the channel W~+~\bbbar,\cite{D0_Higgs}
as shown in Fig.~\ref{D0Higgs}.

\section{Searches for Heavy Gauge Bosons}

New gauge bosons are a feature of many sorts of beyond-the-SM 
physics.     Limits quoted here assume Standard Model
couplings (and in the case of the limit for right-handed $W$ bosons
from \D0,  the same CKM matrix for right and left handed states),
and can be lower for different assumptions.

\subsection{Heavy $W$ Bosons}

A conventional search for a heavy $W$ boson looks for extra events in
the transverse mass plot above the Standard Model W.  Such searches
have low background and the limits are determined mostly by
integrated luminosity (although they do depend on the coupling
assumption, and on the assumption that the accompanying neutrino
is massless).   CDF and \D0 report such 
limits~\cite{CDF_WZprime,D0_Wprime} at 650 and 610 GeV 
respectively.

\D0 has reported two other types of search for a 
heavy $W$ boson.\cite{D0_Wright}
First, there is a search for evidence of a Jacobean peak in the
\Et spectrum of high \Et single inclusive electron events.  This
search would reveal the presence of a heavy W, regardless of 
the handedness of the new particle, and in the limit of a 
massless accompanying neutrino gives a mass limit of 
720~GeV for the heavy $W$ -- significantly higher than 
from either of the transverse mass searches.
This limit
is the solid contour shown in Fig.~\ref{D0Wright}.
The dashed contour in that figure is obtained from a
search for an explicitly right-handed W,  decaying
to an electron plus a right-handed heavy neutrino which in turn
decays to electron plus virtual W, giving as one final
state 2 electrons + 2 jets.  The lack of such events above 
predicted SM background is used to set a limit contour
that reaches higher neutrino masses than the single 
electron search.   This limit does depend on the 
right-handed model used, on the CKM matrix assumption,
and on the mixing between the standard and the new 
state.

\subsection{Heavy $Z$ Bosons}

\D0 and CDF have searched for evidence of a heavy $Z$ boson
in the dilepton invariant mass spectra.  The \D0 
limit ~\cite{D0_Zprime} is
670 GeV, using the dielectron channel only and the 1992-95 
data set.  The CDF limit \cite{CDF_WZprime} is 690 GeV,
using dielectrons and dimuons and the 1992-1995 data set.
Both limits assume standard couplings for the new Z
state.

\section{Searches for Leptoquarks}

Leptoquarks are states of fractional charge which carry both
lepton number and color;
like heavy bosons, they appear in many variations of non-standard
model physics.   The constraints from low energy decays \cite{LQconstraints}
imply that, unless the leptoquark is more massive than the current reach
at the Tevatron, it must have only flavor-conserving couplings.   This
makes the predicted decays simple.  The leptoquarks decay to
leptons and quarks of the same generation only.  The quark quantum
numbers of the leptoquark mean that it can be strongly pair-produced
at the Tevatron, with a cross section independent of the unknown
coupling of the leptoquark to quarks and leptons.

\subsection{First Generation Leptoquarks}

First generation leptoquark pair production will result in three
possible final states:  2 electrons + 2 jets;  1 electron, an
electron neutrino, and 2 jets; or 2 electron neutrinos + 
2 jets.   \D0 has searched for the first two signatures using
1992-93 data.\cite{D0_lq1_1a}   CDF has a result
from its 1989 data for the first signature.\cite{CDF_lq1} 
The latest result is a preliminary limit from \D0 using
1994-95 data, reported in these proceedings.\cite{D0_lq1_1b}
The limit, for the reasonable assumption of equal branching
fraction into electron-quark and neutrino-quark decays, is
$m_{LQ1} > 143$ GeV for a scalar leptoquark.

\subsection{Second Generation Leptoquarks}

Pair production of a second generation leptoquark would 
produce final states analogous to the first generation case,
with electron replaced by muon. Both \D0 and
CDF have results published from the 1992-93 data.\cite{D0_lq2,CDF_lq2_1a}  
CDF also has a preliminary
limit from 1994-95 data \cite{CDF_lq2_1b} of $m_{LQ2} > 141$
GeV 
for a scalar leptoquark decaying with equal branching 
fraction to muon and to muon neutrino.
 
\subsection{Third Generation Leptoquarks}

The third generation case has slightly different phenomenology.
The quarks involved in the leptoquark decays for the first
and second generation leptoquarks would be u,d and c,s
respectively.  But for the third generation, the decays would
be to $\tau$ or $\nu_{\tau}$ plus t or b quarks.  For the
mass range below the top quark,  a leptoquark of charge
state $\pm  {2 \over 3}$ will decay to $\tau + b$ only since
$\nu_{\tau} + t$ is kinematically forbidden.  CDF has 
searched for such a leptoquark,\cite{CDF_lq3} obtaining
a limit of 94 GeV for the scalar case and 220 GeV for the
vector case.

\section{Searches for Fourth Generation Particles}

Another kind of new particle accessible at the Tevatron
would be quarks and leptons from a possible fourth
generation (with massive neutrinos, of course, to 
conform with LEP data).   There are two searches for
such particles presented to date, both from \D0.

\subsection{\bprime\ Quarks}

The charge $1 \over 3$ quark of a fourth generation is
required by LEP I data to be heavier than $1 \over 2$ the
$Z^{0}$ mass, but could possibly be lighter than the 
top quark.   Earlier top searches \cite{D0_toplim} which
did not require b-tagging of the final state would limit
a \bprime\  which decayed semileptonically to charm to a mass
greater than 131 GeV.  However, if the CKM suppression
of the decay to charm is great enough, the \bprime\ might decay
wholly or in part through flavor-changing neutral current
modes, for example $b + \gamma$, $b + gluon$, or
$b + Z$.   For the mass region between the LEP I limit and
$m_{Z} + m_{b} \approx 95$ GeV, the decays to $Z$ are 
forbidden and the branching ratio into $b + \gamma$ has
been calculated.\cite{Bprime_theory}  \D0 has searched for
\bprime\ pair production with two possible final states: $\gamma
+ 3$ jets, one of which is tagged as a $b$ jet, and $ 2 \gamma
+ 2$ jets.  Each search separately rules out a \bprime\ 
decaying via FCNC modes with mass
between 45 and 95 GeV.\cite{D0_bprime}

\subsection{Heavy Neutrinos}

\D0\cite{D0_Heavynulim} has also used its trielectron channel search 
in the 1992-93 data to set
limits on a heavy neutrino that mixes with the electron
neutrino, in terms of the mass and mixing parameter.
The exclusion contour extends the LEP I limit to a mass of
$\approx$ 70 GeV, for mixing parameter $(U_{e4})^{2} > 0.1$.

\section{Searches for New Physics in Distributions}

New physics is not necessarily first signalled by the 
observation of production of a specific new particle type.
It may also appear as a deviation in the shape of a distribution
or as a difference from the expected value of a particular
coupling between particles as calculated within the
Standard Model.   Several limits on new physics are
obtained from analyses of this sort using Tevatron 
data.

\begin{figure*}[htb]
\begin{minipage}{0.48\linewidth}  
\begin{center}
\mbox{\mbox{\epsfig{figure=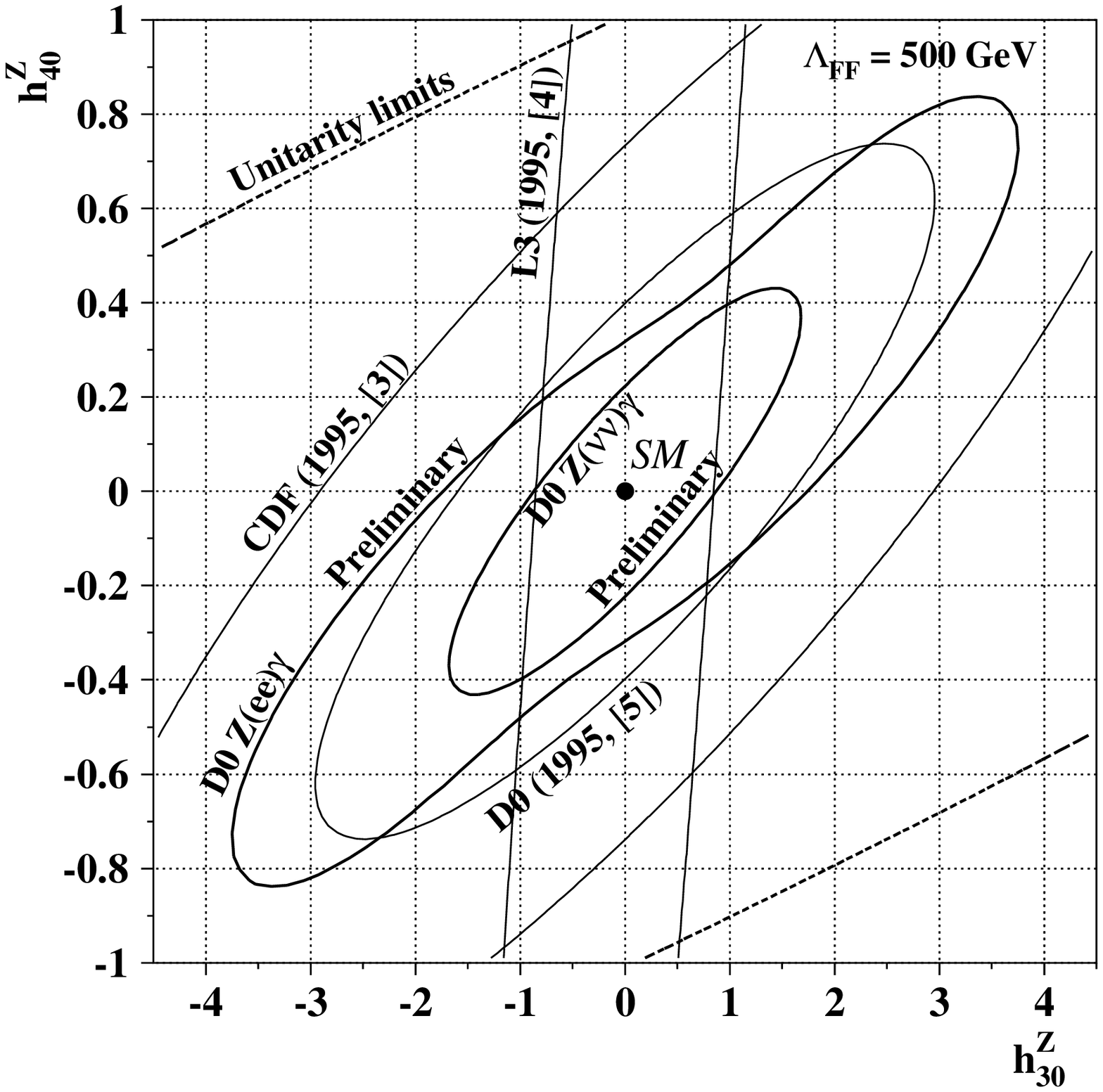,height=60mm}}}
\end{center}
\caption{Anomalous coupling limits on ZZ$\gamma$ channel. }
\label{D0nunugam}
\end{minipage}
\hfill
\begin{minipage}{0.48\linewidth}
\begin{center}
\mbox{\mbox{\epsfig{figure=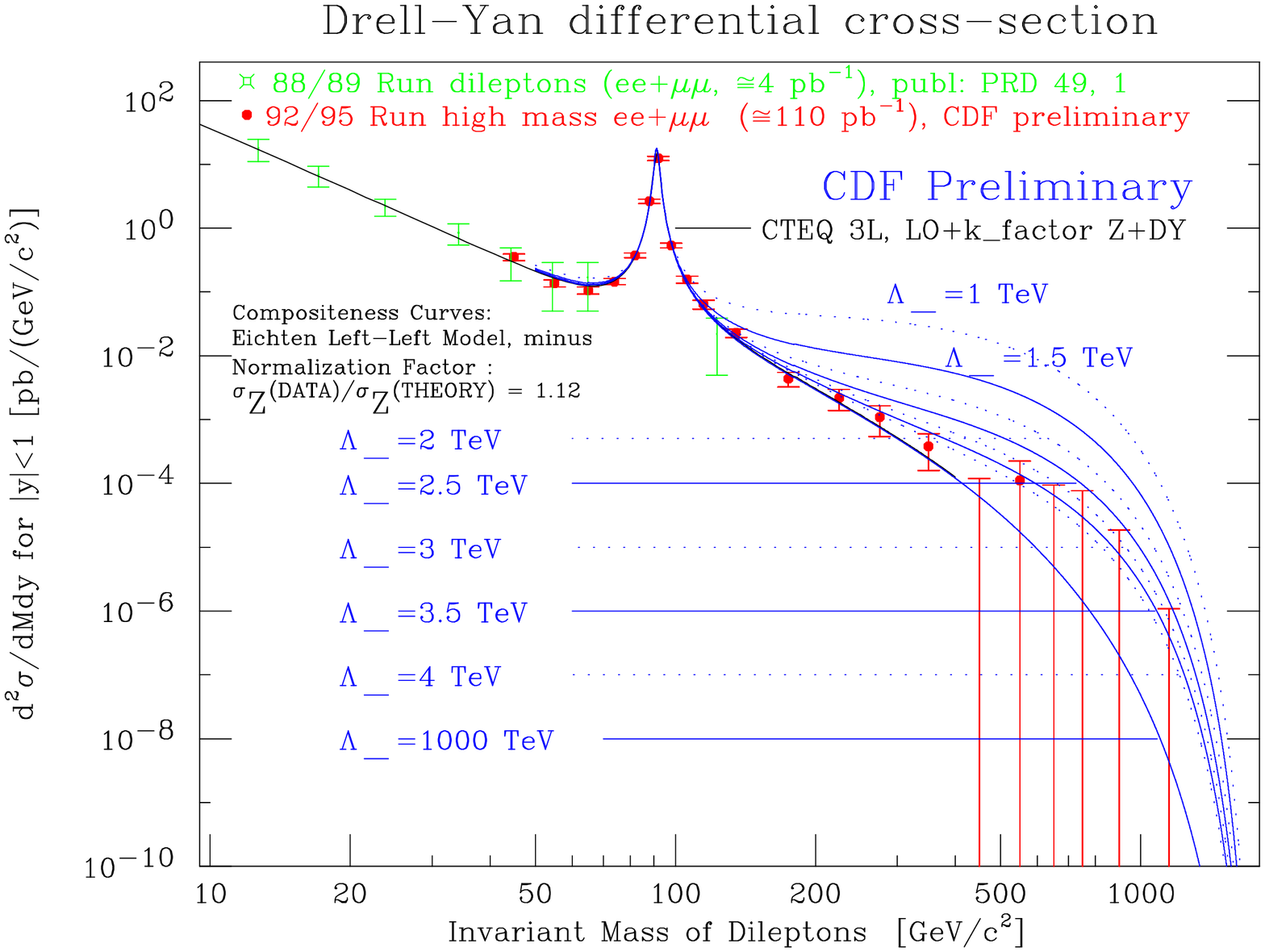,width=50mm}}}
\end{center}
\caption{The spectrum of dilepton invariant mass from CDF, with
fits to various values of the compositeness scale $\Lambda$.}
\label{CDFdrellyan}
\end{minipage}
\end{figure*}

\subsection{Anomalous Triboson Couplings}

One sector which has been diligently searched at the Tevatron
for signs of new phyics is the measurement of triboson
couplings ($WWZ$, $WW\gamma$, $ZZ\gamma$, $Z\gamma\gamma$).
The Standard Model predictions for the couplings are well 
known, and several readily observable final states at the Tevatron
can be used (via counting experiments or fits to $E_{T\gamma}$
spectra) to set limits on new physics expressing itself through
anomalous couplings appearing in these diagrams.  Limits on
these couplings have been set by \D0 \cite{D0_anom,D0_nunugamma}
and CDF 
\cite{CDF_anom} in the $WZ$, $W\gamma$, and $Z\gamma$ final
states, with the $W$ and $Z$ decaying via electron or muon modes.
The tightest limit from such a search is a new report from \D0
\cite{D0_nunugamma} of a measurement in the $Z(\nu\nu)\gamma$
channel, for which the result is shown in Fig. \ref{D0nunugam}.

\subsection{Contact Interactions:  Drell-Yan Mass Distributions}

CDF \cite{CDF_drellyan} has reported limits on the scale 
characterizing a possible contact term interaction between
quarks and leptons, determined by fits to the observed
dilepton invariant mass spectrum.   The spectrum and some of
the fits are shown in Fig. \ref{CDFdrellyan}.  With the
assumption that the contact interaction is between 
left-handed currents and is the same for electrons
and muons, the limits obtained are  $\Lambda_{LL}^{-}
 > 3.8~TeV$ and  $\Lambda_{LL}^{+}
 > 2.9~TeV$ for the cases of constructive and destructive
interference with the SM diagram.   Limits for other assumptions
can be found in Ref. 40.

\begin{figure*}[htb]
\begin{minipage}{0.48\linewidth}  
\begin{center}
\mbox{(a)\mbox{\epsfig{figure=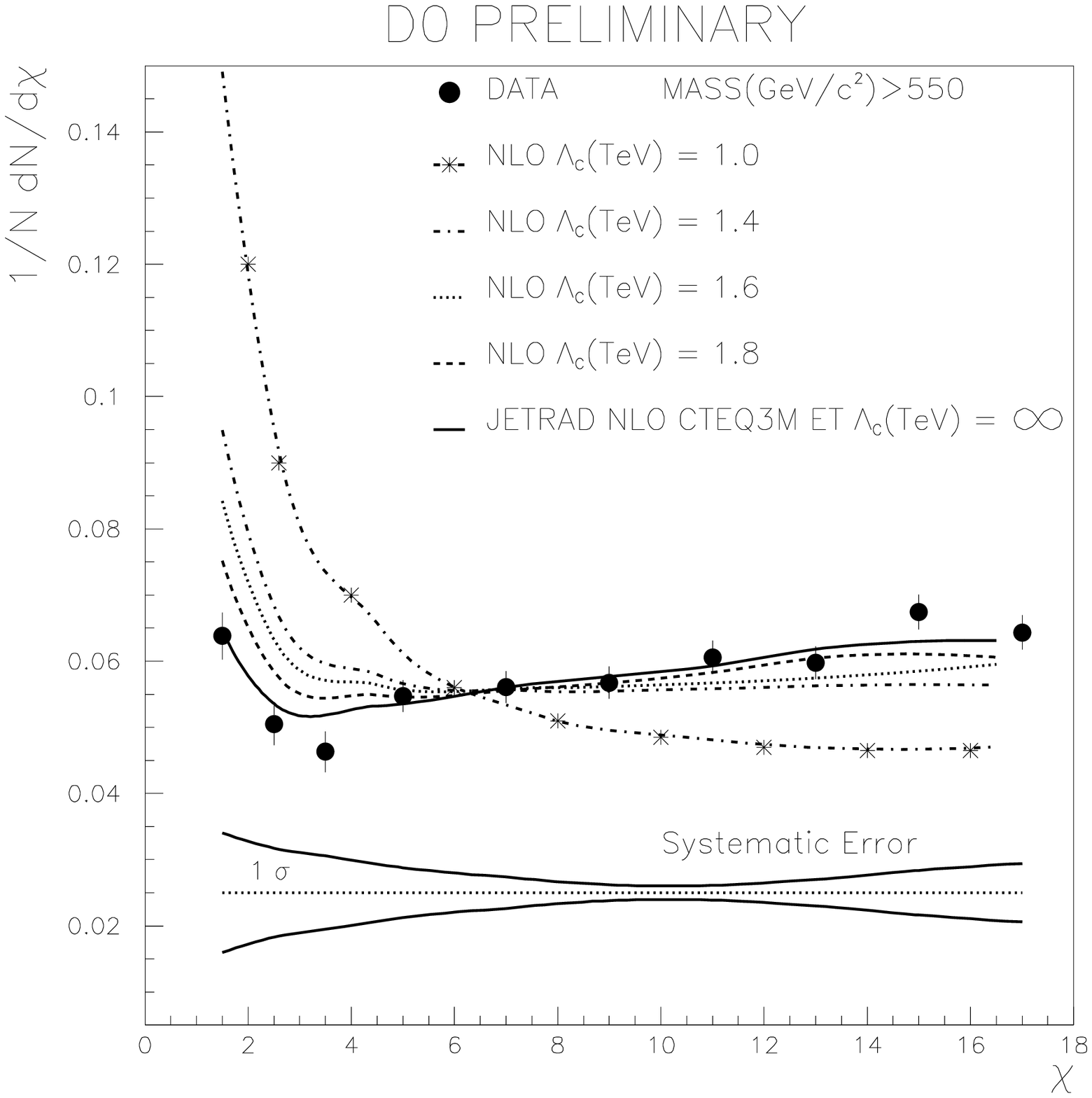,height=60mm}}}
\end{center}
\end{minipage}
\hfill
\begin{minipage}{0.48\linewidth}
\begin{center}
\mbox{(b)\mbox{\epsfig{figure=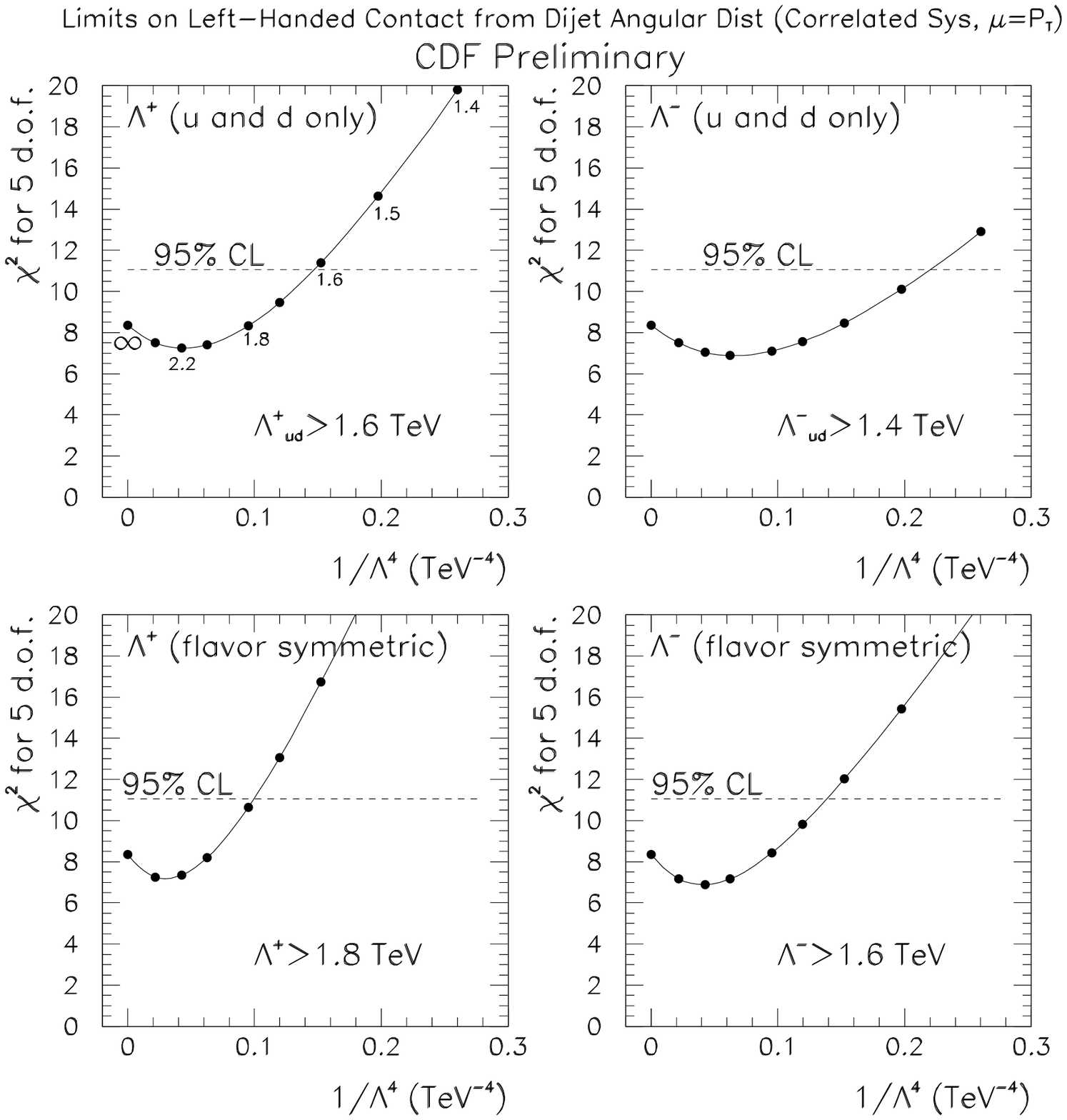,height=60mm}}}
\end{center}
\end{minipage}
\caption{(a) The $\chi$ distribution from \D0, compared to NLO QCD
and to expectations for various compositeness scales $\Lambda$. 
(b) The 95 \%
limits on compositeness scale $\Lambda$, from the ratio of two regions
in $\chi$, from CDF.}
\label{Dijetang}
\end{figure*}

\subsection{Contact Interactions:  Dijet Angular Distributions}

Both \D0 \cite{D0_dijet_ang} and CDF \cite{CDF_dijet_ang} have
reported at this conference on examinations of the angular
distribution of dijet events, searching for evidence of compositeness
in deviations from the NLO predictions of QCD.   Both experiments
use the variable $\chi  = { (1+cos \theta^{*}) \over {(1-cos \theta^{*})}}$  
which is less sensitive to the
assumed parton distribution function than either the cos $\theta$
distribution or the inclusive jet \Et spectrum.  The \D0 result is
shown in Fig. \ref{Dijetang} (a) and the CDF result in Fig.
\ref{Dijetang} (b).

\section{Future Prospects for New Physics at the Tevatron}

This paper has presented a long list of results from the
search for new physics at the Tevatron.  Nevertheless, 
there is still much to be done in this field.  Not all the
analyses presented here include the full 1992-96 
Tevatron data set, nor have all the proposed signatures
for new physics
within the Tevatron's reach been investigated.

\subsection{Further results from Run I}

Supersymmetry limits from the Tevatron have covered most
of the territory of standard, supergravity-inspired models
using the 1992-93 data set.  Several standard channels
still remain to be reported from 1994-96 data.   There
is also the task of more fully exploring the model space:
relaxing the R-parity constraint, for example, or looking 
for less obvious channels such as those suggested in
Ref. 15.   We also need to finish examining the
available leptoquark signatures, to apply the dijet searches
to different new particle cross sections and widths, and
to complete the list of constraints available on new couplings
and contact interaction terms.

\subsection{Expected results from Run II}

In Run II at the Tevatron, the two collider experiments are
expected to accumulate 2 \invfb of integrated luminosity.
In the upgrades to the detectors, CDF will acquire greater
$\eta$ coverage and \D0 will add the capability of 
central track momentum measurement and displaced vertex 
tagging.  Both detectors will thus be able to continue
searching for the signatures explored in Run I, with greater
sensitivity due to the increased luminosity and acceptance.
Table \ref{RunIITable} shows the expected reach for various
new particles in Run II.\cite{TEV2000}  In addition, the
sensitivity to anomalous couplings in the gauge bosons
will increase by a large factor.  New searches
(for example, for \ttbar\  resonances and technicolor)
will become possible.  

\begin{table*}[hbt]
\setlength{\tabcolsep}{1.5pc}
\caption{Expected search reach at the Tevatron in Run II}
\label{RunIITable}
\begin{tabular*}{\textwidth}{lrlr}
\hline
 Particle & Expected mass reach at the Tevatron \\
              &   with 2 \invfb        \\
\hline
        \squark,\gluino          &  $\approx$ 390 GeV \\
    light \stop{1}                &  $\approx$ 150 GeV \\
   \wino{1},\zino{2}             &  $\approx$ 210 GeV \\
   Scalar leptoquarks            &  $\approx$ 240 GeV \\
 (1st and 2nd generation)    &                                    \\
 Heavy $W$ and $Z$                     &  $\approx$ 800-900 GeV \\
  Excited quarks                   &   $\approx$ 800 GeV \\
  Technirhos                         &  $\approx$ 750 GeV \\
\hline
\end{tabular*}
\end{table*}

\section{Conclusion}

Although strenuous efforts have been made to uncover
chinks in its armor, the Standard Model remains so far unscathed 
by Run I searches at  the Tevatron.  
 
 In the search for supersymmetry,  the analyses have begun
to move into the difficult realm of top-like signatures,  as
well as pursuing the traditional \MEt\  channels.  The cleanest
SUSY channel (trileptons) has been investigated with the
full data set from both collider detectors and shows no
hint of a signal.  Some further exploration of the model
dependence of the limits has been undertaken, and has
shown the necessity for careful specification of all
assumptions in interpreting limits.

For other types of new physics, the search has now
been extended to all three generations of leptoquarks
(although we do not have final numbers from both 
experiments).   New gauge bosons and fourth generation
particles
have been searched 
for in novel ways.  Dijet resonance searches have now
been reported from both CDF and \D0,  and also a
prototype $W$-Higgs associated production 
search at \D0.   Constraints on
triboson couplings have been significantly improved,
and limits on contact interaction terms are now being
reported.

A rich program of new physics searches at the Tevatron
will remain interesting into the Main Injector era.

\section*{Acknowledgments}

I would like to acknowledge the organizers of the DPF conference
for a very well-run conference.  Also, I acknowledge help from all
the \D0 and CDF contributors to this conference who submitted
results on new phenomena, as well as the help of Peter Cooper
in supplying the E761 result.

\end{document}